\documentclass{article}
\usepackage[latin9]{inputenc}
\usepackage{amssymb}
\usepackage{esint}

\makeatletter
\newcommand{\lyxaddress}[1]{
\par {\raggedright #1
\vspace{1.4em}
\noindent\par}
}

\date{}

\makeatother

\begin{document}

\title{Classic Calculations of Static Properties of the Nucleons reexamined}

\author{N. F. Nasrallah}

\maketitle

\lyxaddress{\begin{center}
Lebanese University, Faculty of Science, Tripoli, Lebanon
\par\end{center}}
\begin{abstract}
{\normalsize{}Classic calculations of the magnetic moments $\mu_{p}$
and $\mu_{n}$ of the nucleons using the traditional exponential kernel
show instability with respect to variations of the Borel mass as well
as arbitrariness with respect to the choice of the onset of perturbative
QCD. The use of a polynomial kernel, the coefficients of which are
determined by the masses of the nucleon resonances stabilizes the
calculation and provides much better damping of the unknown contribution
of the nucleon continuum. The method is also applied to the evaluation
of the coupling $g_{A}$ of proton to the axial current and to the
strong part of the neutron-proton mass difference $\delta M_{np}$.
All these quantities depend sensitively on the value of the 4-quark
condensate}\\
{\normalsize{}$<0\mid\bar{q}q\bar{q}q\mid0>$ and the value $<0\mid\bar{q}q\bar{q}q\mid0>$
$\simeq$ 1.6$<0\mid\bar{q}q\mid0>^{2}$ reproduces the experimental
results. }\\
\\
Pacs Numbers: 14.20.Dh, 13.40.Em, 13.40.Dk

\pagebreak{}
\end{abstract}

\section{Introduction}

\begin{flushleft}
The QCD sum rule method introduced 35 years ago by Shifman, Vainshtein
and Zakharov \cite{SVZ} has constituted a powerful analytic approach
to the problem of extracting low energy physical quantities from QCD
expressions valid in the space like asymptotic domain.
\par\end{flushleft}

\begin{flushleft}
The method starts from a dispersion integral
\par\end{flushleft}

\begin{flushleft}
\begin{equation}
Residue=\frac{1}{\pi}\int_{th}^{\infty}dte^{-t/M^{2}}ImP(t)\label{eq:1}
\end{equation}

\par\end{flushleft}

\begin{flushleft}
The residue contains the physical quantity of interest and the integral
runs from some physical threshold to infinity. The integral is then
split into two parts
\par\end{flushleft}

\begin{flushleft}
\begin{equation}
\int_{th}^{\infty}dte^{-t/M^{2}}ImP(t)=\int_{th}^{s_{0}}dte^{-t/M^{2}}ImP(t)+\int_{s_{0}}^{\infty}dte^{-t/M^{2}}ImP(t)\label{eq:2}
\end{equation}
Where the divider $s_{0}$ signals the onset of perturbative QCD.
In the first integral on the r.h.s of the equation above ImP(t) describes
the unknown contribution of the resonances. The second integral takes
into account the contribution of the QCD part of the amplitude when
$P(t)$ is replaced by its QCD expression.
\par\end{flushleft}

\begin{flushleft}
$M^{2}$, the square of the Borel mass is a parameter introduced in
order to suppress the unknowns of the problem. If $M^{2}$ is small
the damping of the first unknown integral on the r.h.s of eq.(\ref{eq:2})
is good but the contribution of the unknown higher order non-perturbative
condensates increases rapidly. If $M^{2}$ increases the contribution
of the unknown condensates decreases but the damping in the resonances
region worsens. An intermediate value of $M^{2}$ has to be chosen.
Because $M^{2}$ is an nonphysical parameter the results should be
independent of it in a relatively broad window, this is not the case
in the problems at hand. The choice of the parameter $s_{0}$ which
signals the onset of perturbative QCD is another source of uncertainty.
\par\end{flushleft}

\begin{flushleft}
In this work I shall re-examine the classic calculations of the magnetic
moments of the nucleons \cite{BY,IS} and to the coupling of proton
to the axial current \cite{BK,EPM}, and I shall use polynomial kernels
in dispersion integrals in order to eliminate the contribution of
the unknown integrands. The coefficients of these polynomials are
determined by the masses of the nucleon resonances themselves and
involve none of the instability and arbitrariness inherent to the
use of exponential kernels. The same kernels have been used in refs.\cite{NN}
and \cite{NS} to evaluate the neutron-proton mass difference and
the nucleon mass respectively.
\par\end{flushleft}

\begin{flushleft}
The method will be first applied to the calculation of the magnetic
moments of the nucleons $\mu_{p}$ and $\mu_{n}$ . As a second application
I shall consider the coupling $g_{A}$ of the proton to the axial-vector
current \cite{BK,EPM}, and I shall finally review shortly a previous
calculation of the strong part of the neutron-proton mass difference\cite{NN}.
\par\end{flushleft}

\begin{flushleft}
All the quantities turn out to depend sensitively on the four-quark
condensate 
\par\end{flushleft}

\begin{flushleft}
\begin{equation}
<0\mid\bar{q}q\bar{q}q\mid0>=\kappa<0\mid\bar{q}q\mid0>^{2}\label{eq:3}
\end{equation}
The value of $\kappa$ has been the subject of much investigation
\cite{C} and the result is estimated to vary between 1 (vacuum dominance)
and 4. An interesting result of the present investigation is that
the single value $\kappa$ $\simeq$ 1.6 reproduces the experimental
values of all four quantities $\mu_{p}$, $\mu_{n}$, $g_{A}$ and
$\delta Mnp$.
\par\end{flushleft}

\section{Nucleon Magnetic Moments}

I shall concentrate on the work of Balitsky and Yung \cite{BY} because
it offers better convergence of the OPE than that of Ioffe and Smilga
\cite{IS}

Starting from the 3-point function

\begin{equation}
W_{\mu\nu}(p)=\frac{i}{2}\int\int dxdye^{ip.x}y_{\nu}\langle0\vert Tj_{\mu}(y)\eta(\frac{x}{2})\eta(\frac{-x}{2})\vert0\rangle-\mu\leftrightarrow\nu\label{eq:4}
\end{equation}

where $j_{\mu}$ is the electromagnetic current and 

$\eta(x)=\epsilon_{abc}[u^{a}(x)C\gamma_{\mu}u^{b}(x)]\gamma_{5}\gamma_{\mu}d^{c}(x)$
is the proton current of Ioffe and Smilga \cite{IS}.

The double nucleon pole contribution to the tensor $W_{\mu\nu}(p)$
has the form

\begin{equation}
W_{\mu\nu}(p)=\frac{-\lambda^{2}}{(p^{2}-m_{N}^{2})^{2}}[\frac{i}{2}\{\bar{p},\sigma_{\mu\nu}\}F_{m}^{N}+i\sigma_{\mu\nu}(F_{m}^{N}+F_{e}^{N})+\frac{i}{2}(F_{m}^{N}-F_{e}^{N})\bar{p}\sigma_{\mu\nu}\bar{p}]\label{eq:5}
\end{equation}

where $\bar{p}=p_{\mu}\gamma^{\mu},\,\sigma_{\mu\nu}=\frac{i}{2}[\gamma_{\mu},\gamma_{\nu}]$
and $F_{m}^{N}=2\frac{m_{N}}{e}\mu_{N},F_{e}^{N}=\frac{e_{N}}{e}$
are the values of the magnetic and electric form factors of the nucleons
at zero momentum transfer and $\lambda$ denotes the coupling of the
nucleon current to the nucleon

\begin{equation}
\left\langle 0\left\vert \eta\right\vert N\right\rangle =\lambda U_{N}\label{eq:6}
\end{equation}

In general
\begin{equation}
W_{\mu\nu}(p)=-i\{\bar{p},\sigma_{\mu\nu}\}W_{1}(p^{2})-i\sigma_{\mu\nu}W_{2}(p^{2})-ip\sigma_{\mu\nu}\bar{p}W_{3}(p^{2})\label{eq:7}
\end{equation}

$W_{1}(p^{2})$ is selected which I call $W(p^{2})$ for simplicity.

$W(p^{2})$ has first to be evaluated in the deep euclidean region.
This is carefuly done by Balitsky and Yung \cite{BY} who exploit
both local and bilocal representations of the OPE and who use Vector
Meson Dominance

to estimate the bilocal contributions. The result is
\begin{equation}
W^{QCD}(t)=W_{pert.}(t)+\frac{c_{1}}{t}+\frac{c_{2}}{t^{2}}+\frac{c_{3}}{t^{3}}+...\label{eq:8}
\end{equation}

with
\begin{eqnarray}
c_{1} & = & -\frac{4}{3}\frac{e_{u}}{m_{V}^{2}}\left\langle 0\left\vert \overline{q}q\overline{q}q\right\vert 0\right\rangle \nonumber \\
c_{2} & = & -[\frac{8}{27}\frac{e_{u}}{m_{V}^{2}}\left\langle 0\left\vert \overline{q}q\right\vert 0\right\rangle \left\langle 0\left\vert \overline{u}\sigma Gu\right\vert 0\right\rangle +\frac{1}{3}(e_{d}+\frac{2}{3}e_{u})\left\langle 0\left\vert \overline{q}q\overline{q}q\right\vert 0\right\rangle ]\nonumber \\
 & \simeq & -\frac{1}{3}[\frac{8}{9}e_{u}\frac{m_{0}^{2}}{m_{V}^{2}}+(e_{d}+\frac{2}{3}e_{u})]\left\langle 0\left\vert \overline{q}q\overline{q}q\right\vert 0\right\rangle \label{eq:9}
\end{eqnarray}

Where $m_{V}=m_{\rho}\simeq m_{\omega}$ and $m_{0}^{2}=\frac{\left\langle 0\left\vert \overline{u}\sigma G\right\vert 0\right\rangle }{\left\langle 0\left\vert \overline{q}q\right\vert 0\right\rangle }\backsimeq.8GeV^{2}$
and $c_{3}$ is an unknown term which will be used to estimate the
error.

For small and moderate momentum transfers $W(t=p^{2})$ has double
and single poles
\begin{equation}
W(t)=\frac{\lambda^{2}F_{m}^{p}(t)}{(t-m_{N}^{2})^{2}}+\frac{b_{1}}{(t-m_{N}^{2})}+...\label{eq:10}
\end{equation}

The single pole arises from the unknown nucleon-continuum transitions
and the remainder from the continuum-continuum intermediate states.

As a function of $t$, $W(t)$ is analytic in the complex $t$ plane
with poles shown in eq. (\ref{eq:10}) and a cut along the positive
$t$ axis starting at $t_{th}=(m_{N}+m_{\pi})^{2}.$ 

Consider the contour $C$ consisting of two straight lines just above
and below the cut and running from threshold\ to a large value $R$
and a circle of radius $R$ and consider the integral $\int_{C}dt(m_{N}^{2}-t)f(t)W(t)$

The factor $(m_{N}^{2}-t)$ has been introduced in order to eliminate
the unknown single pole contribution and $f(t)$ is an entire function.
On the circle $W(t)$ can be replaced by $W^{QCD}(t)$ to a good approximation
except possibly near the real axis. Repeated application of Cauchy's
theorem lead to
\begin{eqnarray}
-\frac{1}{2}\lambda^{2}F_{m}^{N}f(m_{N}^{2}) & = & \frac{1}{\pi}\int_{th}^{R}dt(m_{N}^{2}-t)f(t)ImW(t)\label{eq:11}\\
 &  & -\frac{1}{2\pi i}\int_{0}^{R}dt(m_{N}^{2}-t)f(t)DiscW_{pert}(t)+m_{N}^{2}c_{1}-(1+a_{1}m_{N}^{2})c_{2}\nonumber 
\end{eqnarray}

here
\begin{equation}
\frac{1}{2\pi i}DiscW_{p}(t)=\frac{e_{u}}{16\pi^{4}}t\label{eq:12}
\end{equation}
to lowest order in $a_{s}$.

The second term on the r.h.s. of eq. (\ref{eq:11}) equals the contribution
of the integral on the circle of $W_{pert}^{QCD}(t)$. The last two
terms represent the contribution of the integral on the circle of
the first two terms of the non-perturbative expansion of $W^{QCD}(t)$
(for the choice I shall adopt $f(t)=1-a_{1}t-a_{2}t^{2}$) and the
small contribution of the unknown next two non-perturbative terms
has been neglected (note that the use of the exponential kernel would
introduce an infinite number of such unknown in the game).

The first term on the r.h.s. of eq. (\ref{eq:11}) which represents
the contribution of the physical continuum constitutes the main uncertainty
of the calculation. The choice of the so far arbitrary function $f(t)$
aims at reducing this term as much as possible in order to allow its
neglect. The commonly used choice is $f(t)=e^{-t/M^{2}}$ where $M^{2}$
is the Borel mass parameter and it is hoped that the result is not
too sensitive to it. This is not the case in the problem at hand.

I will choose instead a simple polynomial
\begin{equation}
f(t)=p_{0}(t)=1-a_{1}t-a_{2}t^{2}\label{eq:13}
\end{equation}

The coefficients $a_{1,2}$ are chosen in order to minimize $f(t)$
on the interval $I:2GeV^{2}\preccurlyeq t$ $\preccurlyeq3GeV^{2}$
where the nucleon $\frac{1}{2}^{+}$ and $\frac{1}{2}^{-}$ resonances
lie. Minimizing $\int_{I}dtf(t)^{2}$ yields for example
\begin{equation}
a_{1}=.807GeV^{-2},\,a_{2}=-.160GeV^{-4}\label{eq:14}
\end{equation}

with this choice the relative damping $p_{0}(t)/p_{0}(m_{N}^{2})$
does not exceed $6\%$ on the interval $I$. Then
\begin{equation}
\frac{1}{2}\lambda^{2}F_{m}^{p}p_{0}(m_{N}^{2})=\frac{e_{u}}{16\pi^{4}}\int_{0}^{R}dtt(m_{N}^{2}-t)p_{0}(t)-m_{N}^{2}c_{1}+(1+a_{1}m_{N}^{2})c_{2}\label{eq:15}
\end{equation}

Where the contribution of the integral over the resonance region (the
first term on the r.h.s. of eq. (\ref{eq:11})) has been neglected
because of the damping polynomial. The choice of $R$ is determined
by stability considerations, it should not be too small as this would
invalidate the OPE on the circle nor should it be too large because
$p_{0}(t)$ would start enhancing the contribution of the continuum
instead of suppressing it. It turns out indeed that the integral on
the r.h.s. of eq. (\ref{eq:15}) is stable for $2GeV^{2}\lesssim R\lesssim3GeV^{2}$
and that it contributes little compared to the non-perturbative terms
shown in eq.(\ref{eq:9}). The result is essentially proportional
to the four- quark condensate. Numerically then
\begin{equation}
\frac{1}{2}\lambda^{2}F_{m}^{p}f(m_{N}^{2})\backsimeq.80\left\langle 0\left\vert \overline{q}q\overline{q}q\right\vert 0\right\rangle \label{eq:16}
\end{equation}

The coupling $\lambda$ has been obtained by a similar method \cite{NN,NS}
\begin{equation}
(2\pi)^{4}\lambda^{2}m_{N}p_{0}(m_{N}^{2})=-B_{3}I_{1}(R)-B_{7}+a_{1}B_{9}\label{eq:17}
\end{equation}

with
\begin{eqnarray}
B_{3} & = & 4\pi^{2}(1+\frac{3}{2}a_{s})\left\langle 0\left\vert \overline{q}q\right\vert 0\right\rangle \nonumber \\
B_{7} & = & -\frac{4}{3}\pi^{4}\left\langle 0\left\vert \overline{q}q\right\vert 0\right\rangle \left\langle 0\left\vert a_{s}\bar{G}G\right\vert 0\right\rangle \nonumber \\
B_{9} & = & -(2\pi)^{6}\frac{136}{81}a_{s}\left\langle 0\left\vert (\overline{q}q)^{3}\right\vert 0\right\rangle \label{eq:18}\\
I_{1}(R) & = & \int_{0}^{R}dttp_{0}(t)\nonumber 
\end{eqnarray}

Numerically 
\begin{equation}
\lambda^{2}p_{0}(m_{N}^{2})=\frac{1.21GeV^{6}}{32\pi^{4}}\label{eq:19}
\end{equation}

So that with 
\begin{equation}
\left\langle 0\left\vert \overline{q}q\overline{q}q\right\vert 0\right\rangle =\kappa\left\langle 0\left\vert \overline{q}q\right\vert 0\right\rangle ^{2}\label{eq:20}
\end{equation}

I finally get 
\begin{equation}
F_{m}^{p}=1.50\kappa\label{eq:21}
\end{equation}

At this point it is interesting to confront my approach to the use
of the exponential kernel. One would have instead of eq. (\ref{eq:15})
\begin{equation}
\lambda^{2}F_{m}^{p}e^{-\frac{m_{N}^{2}}{M^{2}}}=\frac{e_{u}}{16\pi^{4}}\int_{0}^{R}dtt(m_{N}^{2}-t)e^{-\frac{t}{M^{2}}}-m_{N}^{2}c_{1}+(1-\frac{m_{N}^{2}}{M^{2}})c_{2}\label{eq:22}
\end{equation}

and
\begin{equation}
(2\pi)^{4}\lambda^{2}e^{-\frac{m_{N}^{2}}{M^{2}}}=-B_{3}M^{4}E_{1}(R/M^{2})-B_{7}+\frac{1}{M^{2}}B_{9}\label{eq:23}
\end{equation}

where
\[
E_{1}=\int_{0}^{R/M^{2}}dx\,x\,e^{-x}
\]

The disadvantage of the standard approach is that the result is very
sensitive to the value of the Borel mass parameter $M^{2}$. It actually
varies by a factor of 2.8 when $M^{2}$ varies between $.8GeV^{2}$
and $1.4GeV^{2}.$

The use of the polynomial kernel has stabilized the calculation. The
magnetic moment of the neutron is obtained by the exchange $u\leftrightarrow d$
with the result
\begin{equation}
F_{m}^{n}=-1.26\kappa\label{eq:24}
\end{equation}

The method can likewise be applied to the calculation of the proton
to the axial-vector current to which I turn to in the next paragraph.

\section{Coupling of the Proton to the Axial-Vector Current}

Following Belyaev and Kogan \cite{BK} start with the polarization
amplitude in an external axial-vector field
\begin{equation}
\Pi^{A}(q^{2})=i\int dxe^{iqx}\left\langle 0\left\vert T\eta(x)\eta(0)\right\vert 0\right\rangle _{A}\label{eq:25}
\end{equation}

$\Pi^{A}(q^{2})$ also has double and single nucleon poles
\begin{equation}
\Pi^{A}(t)=-\frac{\bar{\lambda}^{2}g_{A}}{(t-m_{N}^{2})^{2}}+\frac{b\bar{\lambda}^{2}}{(t-m_{N}^{2})}+...\label{eq:26}
\end{equation}

and in the deep Euclidean region
\begin{equation}
\Pi_{QCD}^{A}(t)=t\ln(-t)+\frac{c_{1}}{t}-\frac{20}{9}a_{qq}^{2}.\frac{1}{t^{2}}+\frac{c_{3}}{t^{3}}+...\label{eq:27}
\end{equation}

where
\begin{eqnarray}
c_{1} & = & \frac{1}{4}\left\langle 0\left\vert g_{s}\bar{G}G\right\vert 0\right\rangle +\frac{16}{9}\pi^{2}m_{1}^{2}f_{\pi}^{2}\label{eq:28}\\
c_{3} & = & -\frac{7}{6}m_{0}^{2}a_{qq}^{2}\nonumber 
\end{eqnarray}

with $m_{1}^{2}=1.5GeV^{2},\,\bar{\lambda}^{2}=2(2\pi^{4})\lambda^{2},\,a_{qq}^{2}=4\pi^{2}\left\langle \bar{q}q\right\rangle ^{2}$

The contibution of the single pole has been estimated to be quite
small \cite{BK,IS}. I shall neglect this contribution in order not
to eliminate it by multiplying by $(t-m_{N}^{2})$ as before because
this would introduce higher order unknown condensates whose contribution
could be large because the convergence of the asymptotic series is
not good enough.

The method used in the preceding paragraph is repeated and the same
damping polynomial $p_{0}(t)$ is used which gives
\begin{equation}
-\bar{\lambda}^{2}p_{0}^{^{\prime}}(m_{N}^{2})g_{A}=\int_{0}^{R}dttp_{0}(t)+c_{1}+a_{1}\frac{20}{9}a_{qq}^{2}-a_{2}c_{3}\label{eq:29}
\end{equation}

The final numerical result comes out
\begin{equation}
g_{A}=.39+.61\kappa\label{eq:30}
\end{equation}

\section{Results and Conclusions}

The results obtained for the magnetic moments and the axial-vector
coupling of the nucleon
\begin{eqnarray}
F_{m}^{p} & = & 1.50\kappa,\label{eq:31}\\
F_{p}^{n} & = & -1.26.\kappa,\,g_{A}=.39+.61\kappa\nonumber 
\end{eqnarray}

are very sensitive to the value of $\kappa$ on which no consensus
exists. 

Another physical quantity which is also very sensitive to this parameter
is the strong part of the proton-neutron mass difference
\begin{equation}
\delta M_{np}=(m_{d}-m_{u}).U=(2.60\pm.50)MeV\label{eq:32}
\end{equation}

\begin{flushleft}
This quantity was evaluated in \cite{NN} with the same approach used
here with the result 
\begin{equation}
U=1.03\kappa-.57\label{eq:33}
\end{equation}
and the values taken for the quark masses are $m_{u}=(2.9\pm.2)MeV)$
and $m_{d}=(5.3\pm.4)MeV$ \cite{D} The expression obtained in \cite{NN}
differs slightly from eq. (\ref{eq:32}) this is due to the fact that
the condensate $B_{9}$ appearing in eq. (\ref{eq:18}) was taken
with the wrong sign.
\par\end{flushleft}

\begin{flushleft}
Of course our results are affected by the uncertainties of the calculation.
It is interesting however that a single value of $\kappa$ reproduces
the experimental values of the nucleon magnetic moments, the coupling
to the axial-vector current and the neutron-proton mass difference.
Indeed, taking $\kappa$ = 1.6 eqs. (\ref{eq:20}), (\ref{eq:23}),
(\ref{eq:29}) and (\ref{eq:32}). 
\begin{equation}
F_{m}^{p}=2.40,\,F_{m}^{n}=-2.01,\,g_{A}=1.36,\,\delta M_{np}=2.6MeV\label{(34)}
\end{equation}
which compare well with the experimental numbers 
\begin{equation}
F_{m}^{p}=2.71,\,F_{m}^{n}=-1.91,\,g_{A}=1.27,\,\delta M_{np}=(2.6\pm.5)MeV\label{(35)}
\end{equation}
\pagebreak{}
\par\end{flushleft}

\end{document}